\documentclass[manuscript,screen]{acmart}
\usepackage{makecell}
\usepackage{multirow}

\AtBeginDocument{%
  \providecommand\BibTeX{{%
    \normalfont B\kern-0.5em{\scshape i\kern-0.25em b}\kern-0.8em\TeX}}}

\acmConference[AI Safety and Robustness in Finance Workshop on ICAIF]{Workshop on AI Safety and Robustness In Finance}{
  2023}{New York, NY}
\acmBooktitle{AISRF '23: ICAIF Workshop on AI Safety and Robustness in Finance,
 November 27, 2023, New York, NY} 

\begin{document}

\title{Benchmarking Large Language Model Volatility}


\author{Boyang Yu}
\authornotemark[1]
\email{boy.yu@nyu.edu}
\affiliation{%
  \institution{Center for Data Science, New York University}
  \streetaddress{60 Fifth Ave}
  \city{New York}
  \state{New York}
  \country{USA}
  \postcode{10011}
}

\renewcommand{\shortauthors}{B. Yu}

\begin{abstract}
  The impact of non-deterministic outputs from Large Language Models (LLMs) is not well examined for financial text understanding tasks. Through a compelling case study on investing in the US equity market via news sentiment analysis, we uncover substantial variability in sentence-level sentiment classification results, underscoring the innate volatility of LLM outputs. These uncertainties cascade downstream, leading to more significant variations in portfolio construction and return. While tweaking the temperature parameter in the language model decoder presents a potential remedy, it comes at the expense of stifled creativity. Similarly, while ensembling multiple outputs mitigates the effect of volatile outputs, it demands a notable computational investment. This work furnishes practitioners with invaluable insights for adeptly navigating uncertainty in the integration of LLMs into financial decision-making, particularly in scenarios dictated by non-deterministic information.
\end{abstract}

\keywords{Large language model, Uncertainty quantification, Robustness, Sentiment analysis}


\received[accepted]{19 November 2009}

\maketitle

\section{Introduction}

Large Language Models (LLMs) have demonstrated impressive capabilities in various natural language processing tasks\cite{woodhouse2023can,lopez2023can, lin2023generating}, including sentiment prediction. However, their application in the financial domain presents unique challenges, primarily stemming from the inherently abstract nature of the task\cite{liu2021deep}. 

In the context of financial decision-making, the ability to quantify volatility has paramount importance. Understanding and managing volatility allows for a more informed and robust approach towards financial strategies, whether it's determining the allocation of assets in a portfolio or evaluating the potential risks and returns associated with specific investment opportunities. 

The broader landscape of uncertainty quantification in textual analysis offers crucial insights. In various domains, subjective judgments, vague language, and diverse interpretations necessitate a nuanced approach to uncertainty assessment. For instance, consider sentiment analysis in customer reviews for a product. A statement like "The product is good, but not great" introduces ambiguity. Similarly, in legal documents, the interpretation of contractual clauses may hinge on subtle linguistic nuances. Here, techniques like semantic ambiguity detection or probabilistic modeling provide avenues for handling uncertainty\cite{lin2023generating, liu2021deep}. There are many quantitative metrics to evaluate the level of uncertainty, such as model confidence or entropy\cite{gal2016dropout, lakshminarayanan2017simple}. 

However, traditional uncertainty quantification methods often rely on model generated probabilities or embeddings, requiring direct access to language model's intermediate output. In modern large language models, such as Chat Generative Pre-trained Transformers (ChatGPT)\cite{openai2023gpt4} and Large Language Model Meta AI (LLaMA)\cite{touvron2023llama}, end users often lack a direct access to anything other than the generated text. This lack of transparency can make it challenging to understand how and why the model arrives at a particular prediction or output. As a result, it becomes crucial to develop methods that allow for a practical assessment of uncertainty in order to enhance the reliability and trustworthiness of model outputs.  

The task of news sentiment analysis presents an ideal evaluation ground for assess the level of variation, or volatility. Inherent uncertainties and non-deterministic outputs both contribute to volatility. Furthermore, tracing the effects of volatile outputs in the subsequent investment process provides a tangible demonstration of how these nuances reverberate through each stage of decision-making. A previous study investigates stock price prediction using ChatGPT\cite{lopez2023can}. Researchers extract news sentiment related to price change, aggregate sentiment scores at a daily ticker level, analyze the relationship between sentiment score and excess return, and ultimately construct a portfolio for the US stock market. Their primary focus was on assessing the impact of increased language model capacity on return predictability. However, the volatility perspective remained unexplored. Our work aims to delve deeper, evaluating how volatile model outputs potentially impact investment returns and to what extent.

This study seeks to address a series of pertinent research questions that arise from the interaction between LLMs, financial sentiment prediction, and volatility quantification. We are interested in 

\begin{itemize}
    \item Quantifying LLM uncertainty without direct access to model generated probabilities
    \item Understanding the limitations of volatile sentiment predictions on trading signals execution
\end{itemize}

We first demonstrate that the presence of uncertainty in sentiment classification regardless of temperature setting and its profound implications on downstream tasks, particularly in portfolio construction and return. The resulting variability underscores the need for careful consideration when utilizing LLMs in financial decision-making processes. This discovery reinforces the importance of acknowledging and managing the inherent non-determinism in LLM outputs.

The paper is organized as follows: Section 2 provides background on uncertainty quantification and sentiment analysis for trading strategy. Section 3 covers methodology, including data description and modeling. Section 4 outlines experiment design and results. Finally, Section 5 concludes and discusses key findings. 

\section{Background}

In this section, we describe the specific type of uncertainty under examination in the context of using Large Language Models to perform sentiment analysis for investment decision making, meanwhile introducing terminologies used in the rest of the paper.

\subsection{Pre-trained Large Language Models}

Pre-trained Large Language Models(LLMs) are powered by deep learning techniques and undergo training on extensive text corpora. Their primary objective is to predict the next token given previous sequence. By learning conditional probability distributions for tokens based on prior sequences, LLMs gain a profound understanding of language nuances. 

The success of large language models hinges on their information encoding and token sampling processes. State-of-the-art model architectures, such as GPT and LLaMA, employ transformer architectures with self-attention mechanisms\cite{vaswani2017attention}, enabling them to grasp intricate language nuances. During the sampling phase, tokens with higher probabilities are favored as the next token. This preference is modulated by the temperature parameter, which governs the smoothness of probability distribution. Increasing temperature leads to more creative prediction. In practice, a temperature of 0 is chosen for the goal of determinism\cite{lopez2023can, woodhouse2023can, yang2023large}.

AI community have an ongoing debate about model sharing. Some advocate open source, such as Meta LLaMA, promoting that shared tools encourage collaboration, transparency, and accessibility. Some are extremely cautious in open source, such as OpenAI ChatGPT, arguing that unrestricted access to advanced AI models might lead to potential misuse of the technology\cite{openai2023gpt4}. Despite the attitude, both sides have made outstanding progress in natural language understanding tasks, where GPT-3.5( backend of free version of ChatGPT) and LLaMA2 excel in many benchmarks. 

In the analysis, we use both the GPT and LLaMA model families. Specifically, we focus on the light weighted versions of their interactive chatbot backends, GPT-3.5-turbo and LLaMA-7b. We study the models in a zero-shot setup, i.e., without providing additional training where we present text to models and ask them to provide a response containing sentiment labels. 

\subsection{Sentiment Analysis and Investment Decision-Making}

Sentiment extraction benefits from LLMs' linguistic capabilities, while not explicitly designed for predicting asset prices\cite{lopez2023can, yang2023large, woodhouse2023can}. These models' ability to discern contextual meanings and linguistic patterns could potentially extract valuable insights about a firm's prospects from textual data, such as news headlines, even in the absence of direct financial training. This opens up new avenues for utilizing natural language processing in financial analysis.

Recent studies that use ChatGPT in investment decision-making process include Lopez-Lira and Tang finding ChatGPT is able to predict stock movement\cite{lopez2023can}, and Yang and Menczer\cite{yang2023large} demonstrating ChatGPT successfully identifies credible news outlets. These studies collectively illustrate the potential of Large Language Models (LLMs) in enhancing financial decision-making.

\subsection{Volatility Quantification}

An unspoken rule in combining LLMs and financial text is setting temperature equals 0 to maximize determinism of model output. Now the question is, how much volatility is left and how much will there be if we increase temperature.

Volatility in non-deterministic model outputs refers to the degree of variation or instability in the generated responses of a model. In the context of language models or AI systems, this means that when the same input is given multiple times, the model produces different responses. It's important to understand and manage this volatility based on the specific requirements and goals of the application. 

The temperature parameter plays a pivotal role in regulating the randomness in the generation process. It is a hyper-parameter that controls the likelihood of selecting tokens during text generation. When the temperature is set to a high value, such as 1.0 or 2.0, it introduces more randomness by assigning nearly equal probabilities to a wide range of tokens. In contrast, setting temperature to a low value, such as 0, the generation process becomes less random. 

Published studies favors zero temperature to maximize reproducibility\cite{lin2023generating, yang2023large, lopez2023can, woodhouse2023can}. Note that a temperature of 0 has its downsides. First, it does not guarantee determinism since small numerical discrepancies in floating precision and distributed computation can cause deviations in the output. Moreover, in financial sentiment extraction, zero temperatures limit the model's capability to explore different interpretations and provide a more comprehensive understanding of the text.  

We propose to analyse language processes through their generated samples, because black-box LLMs do not explicitly return probability scores\cite{lin2023generating, liu2021deep}. In light of comparing model output in multiple repetitions in uncertainty quantification literature(MC dropout\cite{gal2016dropout} and deep ensemble\cite{lakshminarayanan2017simple}), we calculate lexical and semantic variation statistics on model outputs across different runs\cite{giulianelli2023comes}. 


\section{Methods}

\subsection{Data}

We utilize two primary data sources for our analysis: Unicorn Data Services (EOD Historica lData | EODHD) for historical price data and news headlines, as well as Aylien for addtional news headlines. The sample period begins in October 2021 (which avoids potential coverage of ChatGPT’s training data) and ends in September 2023. This sample period ensures that our evaluation is based on information not present in the ChatGPT’s training data, allowing for a more accurate “out-of-sample” assessment of its predictive capabilities. For LLaMA, as it's training data is as recent as July 2023, we believe if we keep the strict “out-of-sample” assessment requirement, we will have an extremely short period of evaluation data. So we keep the entire data since October 2021, and are interested in if additional training data in LLaMA improves sentiment analysis and investment performance.

The EODHD provides API to retrieve daily stock prices, and finance news feeds from Yahoo. Our analysis focus on S\&P 500 companies for simplicity and best reproducibility. We filter news feeds mentioning related s\&p tickers and only related to less than three tickers. The Aylien API\cite{hokamp2023news} serves as an additional data source for collecting news headlines from other sources, namely MSN, CNN, Seeking Alpha and Fox Business. In this data source, we explicitly set ticker prominence greater than or equal to 0.8 to ensure only relevance. Furthermore, we eliminate duplicated news entries by headlines and timestamp. Altogether we collect 216,737 headlines for S\&P 500 tickers.

We map the timing in daily price data and dense news feed data by date. The date of news feed is defined by setting a cutoff time for daily trading at 3pm and setting the news effective date as its next actionable trading date. 

\subsection{Prompt}

Prompt engineering is a crucial step in seeking applicable LLMs responses. A concise and instructive prompt provides context and saves computation time. 

For open source models, such as LLaMA2-7B, we apply a customized prompt, see Tab. \ref{tab:prompt} to each of the headlines. The special tokens at the beginning and near the end are LLaMA2 specific template to create templates for system messages in chatbot-like LLMs. The prompt asks the LLM to produce sentiment labels and associate confidence. 

For close source models like GPT-3.5, we need to consider reduce repeated system messages to overcome query rate limit constraint\cite{arefeen2023leancontext}. We deploy a batch-prompt strategy where we integrate 50 news headlines in one prompt, indexed by numbers, shown in Tab. \ref{tab:prompt}. Similarly, we begin and end the prompt with specific instructions. The final instructions in the bracket is to avoid later sentences omitted due to a hard cutoff in model output length.

\begin{table}
\caption{Prompt design. }
  \label{tab:prompt}
\begin{tabular}{l| l|l}
\toprule
 & \makecell[c]{LLaMA} & \makecell[c]{GPT} \\ 
\hline
Input &\makecell[l]{ <s>[INST] \{\{ You are a helpful assistant who only \\ replies according to instructions. Decide the text's \\ sentiment is positive, neutral, or negative. Indicate\\ your confidence using a float between 0 and 1. \\ Text: \textbf{\{HEADLINE\} }  (instruction) Please  answer \\ in the form of SENTIMENT\_LABEL \\(CONFIDENCE). \}\} [/INST]\\ Ans: }  & \makecell[l]{Decide whether each piece of text's sentiment is \\ positive, neutral, or negative. Indicate your \\ confidence for each piece using a float number\\ between 0 and 1. \\ Text:\\ 1. \textbf{\{HEADLINE\_1\}} \\2. \textbf{\{HEADLINE\_2\}} \\ ... \\ 50. \textbf{\{HEADLINE\_{50}\}} \\ Sentiment for each(label and score only):}  \\
\bottomrule
\end{tabular}
\end{table}

\subsection{Volatility metrics}

Lexical volatility is measured by edit distance between model outputs for each headline. For instance, all zero temperature runs generate altogether three outputs, named by $O_i(i=1,...,k)$. The final edit distance is given by $D_{ij} / L_i + D_{ij} / L_j$ where $D_{ij}$ stands for the edit distance between the i-th and j-th output, the $L$ term is the string length. This aggregation ensure the metric is output length invariant. Aggregating headline level metric gives overall volatility in lexical aspect.

Semantic volatility is explicitly measured at different levels, feed-level and ticker-level. For each output, we extract sentiment labels using rule-based string comparison to search for the appearance of "positive", "neutral" and "negative" and possible negations to derive a feed-level sentiment class label for each headline. The final label is 1 if "positive" or "not negative" presents in the response. It is -1 if "negative" or "not positive" is found. All other ambiguous responses are assigned 0 as a proxy for neural. We split the batch-prompt response by headline indexes given in the prompt and use the same rule to match sentiment labels. At feed-level, we calculate the range of sentiment score across different runs. At ticker-level, if a ticker has multiple feeds per day, we take the average sentiment scores as the ticker-level sentiment score and derive the range of variation. Both feed-level and ticker-level ranges are averaged over the entire corpus to measure overall explicit semantic volatility.

Semantic volatility can be implicit, when using volatile sentiment scores for downstream investment decision. We deploy a simple trading daily trading strategy using ticker-level sentiment scores without fitting. These scores are transformed into stock positions by calculating sentiment score deviation from all tickers' historical rolling mean with a look back window of one month. The deviation accounts cross-sectional recent fluctuations and LLMs' bias towards positivity(\cite{seshadri2022quantifying}). We adopt a long-short strategy to adjust holding positions on a daily basis. The back test period covers October 2021 to September 2023.

\section{Results}

For open source LLaMA, we investigate 4 temperature settings (t = 0, 0.25, 0.5, 1.0, following an experimental work with general language generation \cite{kuhn2023semantic}) with 3 repetitions each. For close source ChatGPT, we only verify our findings with 2 temperature settings (t = 0, 1).

\begin{figure}[H]
\centering
\caption{Volatility persists in LLM output. "Feed" refers to variation in individual LLM response. "Ticker" refers to daily sentiment score for each stock. (Left) Average edit distance over model generated text corpus. (Right) Sentiment score discrepancy is inherent at both feed-level and ticker-level.}
\label{fig-semantic_leixical}
\includegraphics[scale=0.45]{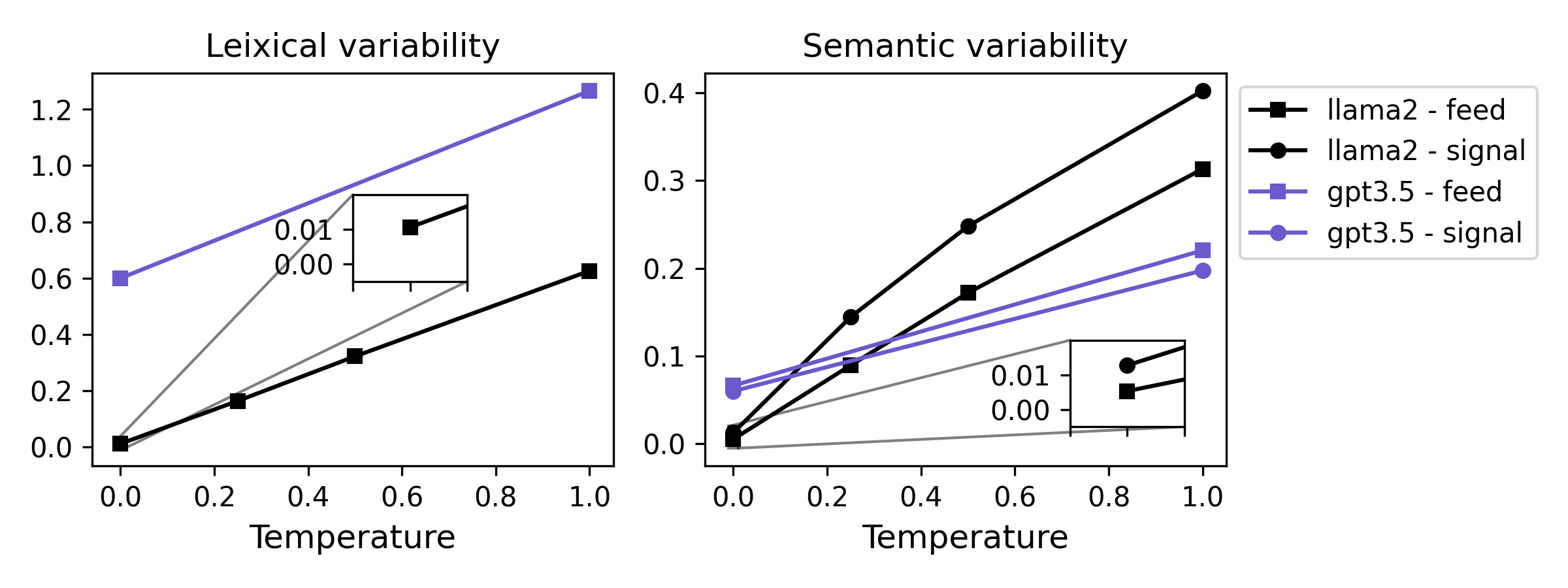}
\end{figure}

\begin{figure}[H]
\caption{A long-short strategy show volatile returns using LLM sentiment scores. The accumulated PnLs are plotted for two settings, low temperature (t=0.0) and high temperature (t=1.0). Each setting is repeated three times. Curves reflect: 1) Non-deterministic model outputs cause instability in portfolio return. 2) High temperature produce more volatile investment strategies.}
\label{fig-pnl}
\centering
\includegraphics[scale=0.55]{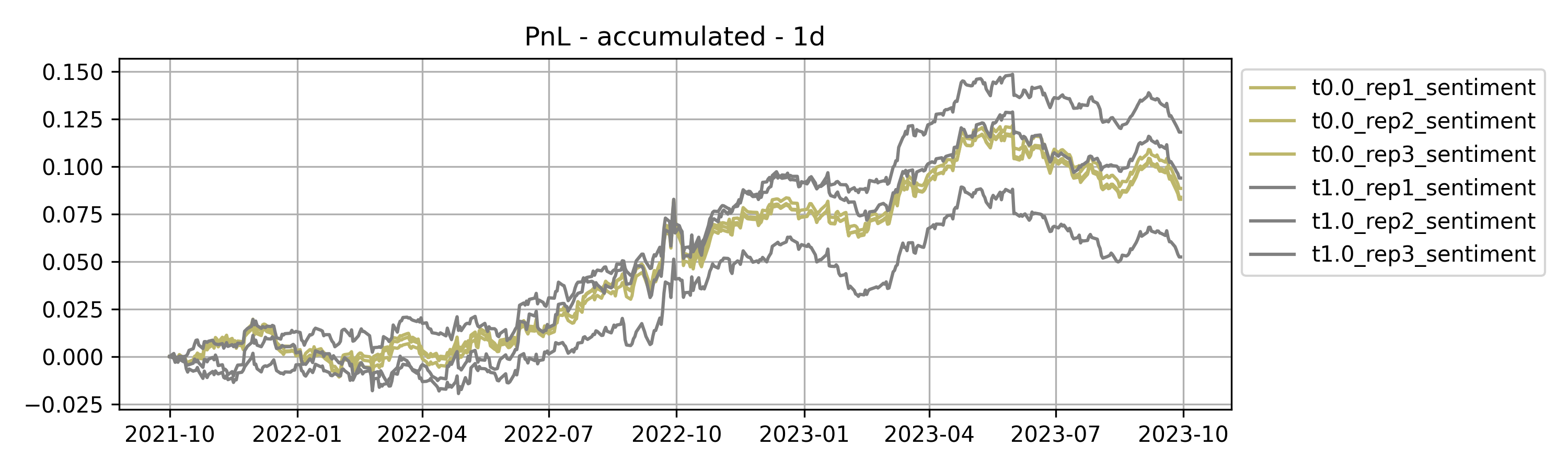}
\end{figure}

\subsection{Lexical and Semantic Volatility Persists in LLM output}

Non-deterministic LLMs outputs show variations in formats and word choices of natural language. Average edit distance between responses are greater than zero for all temperature settings, as shown in Fig. \ref{fig-semantic_leixical}. Notably, even at a zero temperature setting, subtle variations persist. As the temperature increases, the volatility of text generation results also escalates. It it worth noticing that GPT responses are move volatile, partly due to the batch-prompt approach. An interesting future direction is to investigate a low-volatile and token-efficient prompt.

The diversity and variations in the meanings or interpretations of persist in LLM generated responses. In both feed-level and ticker-level, both LLaMA and GPT model show substantial amount of discrepancy across runs. Again, zero temperature does not eliminate semantic volatility. Model-wise, GPT model responses are more stable in changing temperature setting. 

\begin{table}
\begin{center}
\caption{Strategy performance is affected by volatile sentiment scores. We report the mean and standard deviation of total return and Sharpe.}
\label{tab-strat}
\begin{tabular}{ c|c|c|c|c|c  } 
\hline
\multicolumn{2}{c|} {Temperature} & t = 0.0 & t = 0.25 & t = 0.5 & t = 1.0  \\
\hline
\multirow{2}{4em}{LLaMA} & Ret(\%) & 8.51 $\pm$ 0.25 & 8.22 $\pm$ 1.19  & 7.92 $\pm$ 2.64 & 8.82 $\pm$ 2.72 \\ 
& Sharpe & 1.41 $\pm$ 0.04 & 1.48 $\pm$ 0.15 & 1.38 $\pm$ 0.38 & 1.46 $\pm$ 0.45\\ 
\hline
\multirow{2}{4em}{ChatGPT} & Ret(\%) & 6.13 $\pm$ 0.88  & - & - & 7.53 $\pm$ 0.84 \\ 
& Sharpe & 1.12 $\pm$ 0.13 & - & - & 1.45 $\pm$ 0.17\\ 
\hline
\end{tabular}
\end{center}
\end{table}

\subsection{Long-Short Performance Fluctuation with Non-Deterministic Sentiments}

Qualitative volatility in strategy performance is shown in Fig. \ref{fig-pnl}. Lower temperatures are successful in reducing volatility, but they underperform in comparison to the best results achieved at higher temperatures. Notably, the significant disparity (11\% return v.s. 5\% percent return) between the best and worst-performing high-temperature repetitions raises concerns about the robustness of AI in safety-critical systems.

We report quantitative strategy performance in Tab. \ref{tab-strat}. We highlight again the increasing return and Sharpe ratio comes together with more variation in a high temperature setting. Similar to previous subsection, we observe the LLaMA model being more sensitive to the change in temperature. This observation underscores the importance of carefully evaluating the reliability and performance of AI models in the context of automated trading applications.

\bibliographystyle{acm}

\bibliography{sample-base}

\section{Acknowledgments}

The author appreciates insightful discussion with Minhua Zhang and Michal Mucha.

\section{Appendices}








\section{Cost}

EODHD API subscription costs \$19.99. ChatGPT costs about \$14.5 for one pass (estimated error rate about 5\%, need addtional budget to rerun these queries). All ChatGPT experiments costs \$ 67.98.

\end{document}